\documentclass[a4paper, 11pt]{article}
\usepackage{amsmath,amsfonts,amsthm,amssymb,stmaryrd}
\usepackage{graphicx,color}
 \usepackage{bbold}
\usepackage{mathrsfs}
\usepackage{setspace}
\usepackage{tikz}
\usepackage{comment}
\usepackage{appendix}
\usepackage[british]{babel}

\usepackage[unicode]{hyperref}
\hypersetup{colorlinks=true, breaklinks=true, linkcolor=black, menucolor=black, urlcolor=black!20!green!70!blue,citecolor=black!20!green!70!blue}
\usepackage{hyperref}
\usepackage[active]{srcltx}
\usepackage{subfig}
\numberwithin{equation}{section}
\def\l{\lambda}
\def\L{\Lambda}
\def\m{\mu}

\def\S{\Sigma}
\def\t{\tau}

\def\cH{{\cal H}}

\def\cL{{\cal L}}
\def\cM{{\cal M}}

\def\cS{{\cal S}}

\def\cW{{\cal W}}

\def\tr{\text{tr}}
\def\sgn{\mathop{\rm sgn}}

\newcommand{\N}{\mathbb{N}}
\newcommand{\Z}{\mathbb{Z}}

\newcommand{\Hilb}{\mathcal{H}}
\newcommand{\T}{\mathbb{T}}

\newcommand{\order}[1]{\mathcal{O}\!\left( #1 \right)}

\def\12{\frac{1}{2}}

\newcommand\blfootnote[1]{%
  \begingroup
  \renewcommand\thefootnote{}\footnote{#1}%
  \addtocounter{footnote}{-1}%
  \endgroup
}

\begin{document}

\vspace*{-1.5cm}
\thispagestyle{empty}
\begin{flushright}
AEI-2014-034
\end{flushright}
\vspace*{2.5cm}
\begin{center}
{\Large
{\bf The large level limit of Kazama-Suzuki models}}
\vspace{2.5cm}

{\large Stefan Fredenhagen, Cosimo Restuccia}
\blfootnote{{\tt E-mail: FirstName.LastName@aei.mpg.de}}

\vspace*{0.5cm}

Max-Planck-Institut f{\"u}r Gravitationsphysik\\
Albert-Einstein-Institut\\
Am M{\"u}hlenberg 1\\
14476 Golm, Germany\\
\vspace*{3cm}

{\bf Abstract}\\[6mm]
\begin{minipage}{14cm}Limits of families of conformal field theories are of interest in the context of AdS/CFT dualities. 
We explore here the large level limit of the two-dimensional $\mathcal{N}=(2,2)$ superconformal $\mathcal{W}_{n+1}$ minimal models
that appear in the context of the supersymmetric higher-spin AdS$_{3}$/CFT$_{2}$ duality. These models are constructed as Kazama-Suzuki coset models of the form $SU(n+1)/U(n)$. We determine a family of boundary conditions in the limit theories, and use the modular bootstrap to obtain the full bulk spectrum of $\mathcal{N}=2$ super-$\mathcal{W}_{n+1}$ primaries in the theory. We also confirm the identification of this limit theory as the continuous orbifold $\mathbb{C}^{n}/U(n)$ that was discussed recently.
\end{minipage}
\end{center}
\newpage

\tableofcontents
\vspace*{2cm}

\section{Introduction}

Limits of rational conformal field theories in two dimensions play an important role for the proposed higher-spin AdS$_{3}$/CFT$_{2}$ dualities. After the observation that higher-spin gauge theories on asymptotically Anti-de Sitter backgrounds have large asymptotic symmetry $\mathcal{W}$-algebras~\cite{Campoleoni:2010zq,Henneaux:2010xg}, a concrete proposal for the dual of the bosonic Prokushkin-Vasiliev model~\cite{Prokushkin:1998bq} was formulated~\cite{Gaberdiel:2010pz}, which is given by a certain limit of bosonic $\mathcal{W}_{n}$-minimal models. This has been generalised to duality proposals for $\mathcal{N}=2$ \cite{Creutzig:2011fe,Candu:2012jq,Candu:2012tr} and $\mathcal{N}=4$ \cite{Gaberdiel:2013vva} supersymmetric situations. In a very interesting recent development \cite{Gaberdiel:2014cha} it was shown that a certain large level limit of the models occurring in the $\mathcal{N}=4$ case is related to a conjectured conformal field theory dual of tensionless strings on $AdS_{3}\times S^{3}\times T^{4}$, thus pointing towards an understanding how higher-spin gauge theories are related to a tensionless limit of string theory. 

In this article we want to investigate the large level limit of the $\mathcal{N}=2$ superconformal models that appear in the $\mathcal{N}=2$ higher-spin AdS$_{3}$/CFT$_{2}$ duality. These are the Grassmannian Kazama-Suzuki models~\cite{Kazama:1988uz,Kazama:1989qp} that are realised as coset models of the form
\begin{equation}
\frac{\mathfrak{su}(n+1)_k\oplus \mathfrak{so}(2n)_1}
{\mathfrak{su}(n)_{k+1}\oplus\mathfrak{u}(1)_{n(n+1)(k+n+1)}}\ .
\end{equation}
Similarly to what happens for the bosonic models in the large level limit~\cite{Gaberdiel:2011aa}, for $n=1$ it was shown in~\cite{Fredenhagen:2012bw} that the limit theory coincides with the continuous orbifold $\mathbb{C}/U(1)$. Analogously, it was conjectured in~\cite{Restuccia:2013tba} that the large level limit for general $n$ is given by the continuous orbifold $\mathbb{C}^{n}/U(n)$. Recently evidence for this conjecture has been given in~\cite{Gaberdiel:2014vca}, where the untwisted sector as well as the ground states of the twisted sectors of the orbifold theory were identified in the limit theory. In the present work we on the one hand give further support to the conjecture by an analysis of boundary conditions and boundary partition functions, and on the other hand we provide a complete description of the spectrum of $\mathcal{N}=2$ $\mathcal{W}_{n+1}$-primaries in the limit theory, which is based on the modular bootstrap.
\smallskip

The paper is organised as follows. Section~2 contains a short summary of the facts about Kazama-Suzuki models that we will need in this paper. In section~3 we study boundary conditions in Kazama-Suzuki models in the large level limit. We determine the boundary partition functions for discrete boundary conditions in the limit theory, and show that they coincide with the boundary partition functions of fractional boundary conditions in the orbifold $\mathbb{C}^{n}/U(n)$. We present in section~4 a proposal how the full continuous bulk spectrum of $\mathcal{N}=2$ $\mathcal{W}_{n+1}$ primaries arises from the Kazama-Suzuki spectra in the limit. We confirm this proposal by the modular bootstrap that we discuss in section~5.

\section{Kazama-Suzuki Grassmannian cosets}\label{sec:KS-models}

We are interested in the Kazama-Suzuki models~\cite{Kazama:1988uz,Kazama:1989qp}, which are rational $\mathcal{N}=(2,2)$ superconformal field theories based on the coset
\begin{equation}
\frac{\mathfrak{g}_k\oplus\mathfrak{so}(\text{dim} [\mathfrak{g}]-\text{dim} [\mathfrak{h}])_1}{\mathfrak{h}_{k+g_G-g_H}}\ ,
\end{equation}
where $g_G,g_H$ indicate the dual Coxeter numbers of the numerator and denominator algebra, respectively.
A particular class of Kazama-Suzuki models are the Grassmannian cosets, which are specified by two positive integers, the rank $n$ and the level $k$ of the model, and whose explicit coset description reads 
\begin{equation}\label{def-Grassmannian}
\frac{\mathfrak{su}(n+1)_k\oplus \mathfrak{so}(2n)_1}
{\mathfrak{su}(n)_{k+1}\oplus\mathfrak{u}(1)_{\kappa}}\ .
\end{equation}
Here, $\kappa=n(n+1)(k+n+1)$ and the central charge is $c=\frac{3nk}{k+n+1}$. The theories are rational with respect to an extension of the $\mathcal{N}=2$ superconformal algebras, the so-called $\mathcal{N}=2$ $\mathcal{W}_{n+1}$-algebras.

The map of the denominator into the numerator group in equation~\eqref{def-Grassmannian} is specified by the following group homomorphisms (see e.g.\ \cite{Candu:2012jq}):
\begin{align}\label{embedding-H-into-G}
\begin{array}{llll}
i_1&:U(n)\longrightarrow SU(n+1)\ ,\qquad
& i_1 (h,\xi) = \begin{pmatrix}
h\xi & 0\\
0& \xi^{-n}
\end{pmatrix} & \in SU (n+1) \ ,\\[0.45cm]
i_2&:U(n)\longrightarrow SO(2n)\ ,\qquad
& i_{2} (h,\xi) = \begin{pmatrix}
\text{Re} (h\xi^{n+1}) & \text{Im} (h\xi^{n+1})\\
-\text{Im} (h\xi^{n+1}) & \text{Re} (h\xi^{n+1})
\end{pmatrix} & \in SO (2n) \ ,
\end{array}
\end{align}
where $h\in SU (n)$ is a $n\times n$-matrix, and $\xi \in U (1)$ is a phase.

Following the usual coset construction~\cite{Goddard:1985vk,Goddard:1986ee}, the spectrum of the theory is given by the branching of the decomposition of the representations of the numerator algebra in terms of representations of the denominator algebra,
\begin{equation}\label{ch3:KS-hilb-decomposition}
{\cal H}_{\mathfrak{su}(n+1)}^{\Lambda}\otimes {\cal H}_{\mathfrak{so}(2n)}^{\S}=\sum_{\lambda,\mu}
{\cal H}^{\Lambda,\S}_{\l,\m}\otimes \left[{\cal H}^{\l}_{\mathfrak{su}(n)}\otimes {\cal H}^{\m}_{\mathfrak{u}(1)}\right]\ .
\end{equation}
The representations ${\cal H}^{\Lambda,\S}_{\l,\m}$ are labelled by~$(\L,\S;\l,\m)$, where~$\L=(\L_1,\dots,\L_n)$ is a dominant weight of~$\mathfrak{su}(n+1)_{k}$ (where we display only the corresponding weight of the finite-dimensional algebra), $\S$ one of the four dominant weights of~$\mathfrak{so}(2n)_{1}$ ($\S=0$ singlet, $\S=v$ vector, $\S=s$ spinor, $\S=c$ co-spinor), $\l=(\l_1,\dots,\l_{n-1})$ labels dominant weights of~$\mathfrak{su}(n)_{k+1}$, and~$\mu$ is an integer ($\kappa$-periodic) labelling the primaries of the free boson compactified at radius~$\sqrt{\kappa}$.

We are here only interested in the Neveu-Schwarz sector where $\Sigma=0$ or $\Sigma=v$. Also we are usually interested in representations of the full $\mathcal{N}=2$ $\mathcal{W}_{n+1}$-algebra and not only in representations of the bosonic part of it. To get those in the Neveu-Schwarz sector we have to consider in $\mathfrak{so}(2n)_{1}$ the representation $\cH_{\mathfrak{so}(2n)}^{0}\oplus \cH_{\mathfrak{so} (2n)}^{v}$. 

We can get the full $\mathcal{N}=2$ Neveu-Schwarz characters $\Xi^{\L}_{\l,\m}(q)$ of the coset representations by decomposing the product of a $\mathfrak{su}(n+1)_{k}$ character and the character for the $\mathfrak{so}(2n)_{1}$-part, 
\begin{equation}\label{finite-k-decomposition-characters}
\chi^{\mathfrak{su}(n+1),k}_{\L}(q;i_1(h,\xi))\ \theta^{\text{NS}}(q;i_{2}(h,\xi)) =\sum_{\l,\m}\Xi^{\L}_{\l,\m}(q)\ \left[\chi_{\l}^{\mathfrak{su}(n),k+1}(q;h)\ \Theta_{\mu,\kappa}(q;\xi)\right]\ ,
\end{equation}
where $\chi^{\mathfrak{su}(n+1),k}_{\L}$ and $\chi_{\l}^{\mathfrak{su}(n),k+1}$ are the characters of the representations of $\mathfrak{su}(n+1)_k$ and of $\mathfrak{su}(n)_{k+1}$, respectively, $\Xi$ is the coset character, $\Theta$ are $\mathfrak{u}(1)_{\kappa}$ characters; the squared bracket expression on the right hand side is altogether a $\mathfrak{u}(n)$ affine character. $\theta^{\text{NS}}$ is the character of $2n$ Neveu-Schwarz Majorana fermions, given by
\begin{equation}
\begin{split}
\theta^{\text{NS}}(q)= & \chi^{\mathfrak{so}(2n),1}_{0}(q;i_2(h,\xi))+\chi^{\mathfrak{so}(2n),1}_{v}(q;i_2(h,\xi))\\
= & \tr_{\text{NS}} \left( q^{L_{0}-\frac{n}{24}}i_{2}(h,\xi)\right) = 
q^{-\frac{n}{24}}\prod_{j=1}^n\prod_{m=0}^{\infty}\big(1+h_j\xi^{n+1}q^{m+\frac{1}{2}}\big)\big(1+\bar h_j\xi^{-(n+1)}q^{m+\frac{1}{2}}\big)\ .
\end{split}
\end{equation}
Note that $i_{2}(h,\xi)$ given in~\eqref{embedding-H-into-G} has eigenvalues $h_{j}\xi^{n+1}$ and $\bar{h}_{j}\xi^{-n-1}$, where $h_{j}$ are the eigenvalues of $h\in SU(n)$. 

Being class functions, the characters depend only on the coordinates of the Cartan torus $\T^n$ of $U(n)$, which we parameterise as
\begin{equation}\label{introduction-theta-angles}
\T^n\ni\text{Diag}(e^{i\theta_1},\dots,e^{i\theta_n})\ , \quad\text{with}\quad e^{i\theta_j}=h_j\xi^{n+1}\ . 
\end{equation}
The spectrum obtained in this way is subject to selection rules, since some of the representations of the denominator do not appear as subsectors of the numerator. Only those representations occur for which 
\begin{equation}\label{ch3:KS-selection-rules}
\frac{|\Lambda|}{n+1}-\frac{|\lambda|}{n}+\frac{\mu}{n(n+1)}\ \in\ \Z\  ,
\end{equation}
where for $\mathfrak{su}(n)$ and $\mathfrak{su}(n+1)$ the symbol $|\mathcal{D}|$ denotes the number of boxes of the Young diagram associated with the finite dimensional representation $\mathcal D$.

Furthermore, some states in the spectrum have to be identified, because of the outer automorphisms of the numerator and denominator chiral algebras. The group of outer automorphisms of the affine algebra $ \mathfrak{su}(n)_k$ is isomorphic to the centre of the group $SU(n)$, which is $\Z_{n}$.
The automorphism acts therefore by permuting the Dynkin labels of the representations on both the special unitary numerator and denominator algebras, resulting in a $\Z_{n(n+1)}$ group.
Under the automorphisms the $\mathfrak{u}(1)$ label gets shifted by $k+n+1$.
The identifications among representations are accordingly:
\begin{align}\label{ch3:KS-identifications}
\begin{array}{lrll}
\mathfrak{su}(n+1)_k\ :\ & \L_1&\longrightarrow &k-\sum_{j=1}^{n}\L_j\ ,\quad \L_i\longrightarrow \L_{i-1}\quad\text{for}\quad 2\leq i\leq n\ ,\\[2mm]
\mathfrak{su}(n)_{k+1}\ :\ &\l_1&\longrightarrow &k+1-\sum_{j=1}^{n-1}\l_j\ ,\quad \l_i\longrightarrow \l_{i-1}\quad\text{for}\quad 2\leq i\leq n-1\ ,\\[2mm]
\mathfrak{u}(1)_{n(n+1)(n+k+1)}\ :\ & \mu&\longrightarrow &\mu+k+n+1\mod n(n+1)(n+k+1)\ .
\end{array}
\end{align}

Conformal weights can be determined from the knowledge of the~$L_0$ eigenvalues for the representations in the numerator and in the denominator.
The~$U(1)$-charge can be obtained by a careful definition of the fields which generate the~$\mathcal{N}=2$~superconformal algebra~\cite{Kazama:1988uz}.
The conformal dimension $h$ and the $U(1)$ charge $Q$ of the Neveu-Schwarz representation $(\Lambda;\lambda,\mu)$ are then
\begin{align}\label{ch3:confweight}
h &=\ \frac{1}{k+n+1}\left[C^{(n+1)}(\L)
   -C^{(n)}(\l)
   -\frac{\mu^2}{2n(n+1)}\right]
   +h_{\S} \mod \frac{1}{2}\\[0.2cm]
Q &=\ -\frac{\mu}{n+k+1}+Q_{\S}\mod 1\ ,
\end{align}
where $C^{(n)}$ and $C^{(n+1)}$ denote the quadratic Casimir, which for $\mathfrak{su}(n)$ reads
\begin{equation}\label{defCasimir}
C^{(n)}(\l)=
\sum_{1\leq i<j\leq n-1}\!\!\l_i\l_j\ \frac{i(n-j)}{n}+\frac{1}{2}\sum_{j=1}^{n-1}\l_j^2\ \frac{j(n-j)}{n}+\frac{1}{2}\sum_{j=1}^{n-1}\l_j\ j(n-j)\ .
\end{equation}

\section{Boundary conditions in the large level limit}

In this section we analyse the behaviour of boundary states of the coset theory in the large level limit. We show how one can obtain a certain discrete class of boundary conditions for the limit theory, and we study their boundary partition functions. 
We then give support to the claim that the limit theory can be described by the $\mathcal{N}=(2,2)$ superconformal continuous orbifold~$\mathbb{C}^n/U(n)$.

\subsection{Boundary conditions and spectrum}
The Grassmannian coset models $\mathfrak{su}(n+1)/\mathfrak{u}(n)$ are rational with respect to the $\mathcal{N}=2\ W_{n+1}$-algebras. With the Cardy construction~\cite{Cardy:1989ir} one can define rational boundary states, i.e.\ those that preserve one copy of the chiral algebra of the theory. Depending on the gluing conditions for the supercurrents, one can distinguish A-type and B-type boundary conditions -- here we want to focus on A-type conditions. For a diagonal bulk spectrum, A-type boundary conditions are labelled by the same set of representations as the bulk fields, i.e.\ by tuples $(\L,\S;\l,\m)$, with $\L,\S,\l$ dominant weights of~$\mathfrak{su}(n+1)_{k},\mathfrak{so}(2n)_{1},\mathfrak{su}(n)_{k+1}$ respectively, and~$\m$ the~$U(1)$ label of the free boson on the circle of radius $\sqrt{\kappa}$. We want to use a particular choice for the gluing condition for the fermions, this restricts the $\mathfrak{so}$-label to $\S=0$ or $\S=v$.

From the analysis of boundary renormalisation group flows in coset models~\cite{Fredenhagen:2001kw,Fredenhagen:2002qn,Fredenhagen:2003xf} one knows that a
boundary state with $\mathfrak{su}(n+1)$ label $(\L_1,\dots \L_n)$ can be obtained by a flow from a superposition of boundary states with $\mathfrak{su}(n+1)$ label $(0,\dotsc ,0)$. These boundary renormalisation group flows become shorter and shorter as $k$ grows, and in the limit $k\to\infty$ the initial and final fixed-point coincide. We therefore expect that in the limit theory the elementary boundary conditions are labelled by the tuples~$(0,\cS;\cL,\cM)$. 
This fact generalises what we observed in the case $n=1$ of $\mathcal{N}=2$ minimal models~\cite{Fredenhagen:2012rb,Fredenhagen:2012bw}.

The boundary partition function for two boundary conditions 
$(0,\cS_{i};\cL_{i},\cM_{i})$ ($i=1,2$) in the model at level $k$ is given by
\begin{equation}
Z_{(0,\cS_{1};\cL_{1},\cM_{1}) (0,\cS_{2};\cL_{2},\cM_{2})} (\tilde{\tau}) = 
\sum_{\Sigma,\lambda} N^{\mathfrak{so}(2n)_{1}}_{\Sigma\cS_{1}}{}^{\cS_{2}} N^{\mathfrak{su}(n)_{k+1}}_{\lambda\cL_{1}}{}^{\cL_{2}} \,\chi_{(0,\Sigma;\lambda,\cM_{2}-\cM_{1})} (\tilde{q}) \ ,
\end{equation}
where $\tilde{q}=e^{2\pi i\tilde{\tau}}$, and the symbols $N$ denote the fusion coefficients. In the expression above, the characters of the bosonic subalgebra of the coset algebra appear. To simplify our analysis we want to study the supersymmetric partition functions with the full super-$\cW_{n}$ characters $\Xi^{\Lambda}_{\lambda,\mu}$ (the unprojected partition function). Then we can forget about the $\mathfrak{so}$-label $\cS$ in the boundary conditions, and the unprojected partition function reads
\begin{equation}\label{boundpartfunc}
Z_{(0;\cL_{1},\cM_{1})(0;\cL_{2},\cM_{2})} (\tilde{\tau}) = \sum_{\lambda} N^{\mathfrak{su}(n)_{k+1}}_{\lambda\cL_{1}}{}^{\cL_{2}} \, \Xi^{0}_{\lambda,\cM_{2}-\cM_{1}} (\tilde{q})\ .
\end{equation}
To obtain the large level limit $k$ of this boundary partition function we have to identify the limit of the coset characters, which we will do in the following subsection.

\subsection{Limit of coset characters}

We want to evaluate the coset characters $ \Xi^{0}_{\lambda,\mu}$ in the limit $k\to\infty$ while keeping the labels $\lambda,\mu$ fixed. This has been worked out in~\cite{Candu:2012jq,Gaberdiel:2014vca} (see~\cite{Gaberdiel:2011zw} for a similar analysis for the bosonic models) by using large level expansions of the individual character that enter~\eqref{finite-k-decomposition-characters} as we will review below.

The $\mathfrak{su}(n)_{k+1}$ character is given by the Weyl-Ka\v c formula
\begin{equation}
\chi^{\mathfrak{su}(n)_{k+1}}_{\lambda}=q^{-\frac{(n+1)(n-1)(k+1)}{24(k+n+1)}}\frac{\sum_{w\in\hat W} \sgn(w) e^{w(\lambda+\rho)} }{\sum_{w\in\hat W} \sgn(w)e^{w(\rho)}}
\end{equation}
where $\hat W$ represents the affine Weyl group, which is given by the semidirect product of finite Weyl reflections with translations of elements of the root lattice.
The affine translations contribute with terms of order $q^{k+1-\sum_i\lambda_i}$ (see e.g.~\cite{Fuchs:1997jv}), which are suppressed as $k$ becomes large.
It is therefore possible to write down an expansion of the form
\begin{equation}\label{k-expansion-su(n)-character}
\chi^{\mathfrak{su}(n)_{k+1}}_{\lambda}(q;h)=q^{-\frac{(n+1)(n-1)(k+1)}{24(k+n+1)}+\frac{C^{(n)}(\lambda)}{n+k+1}} \frac{\text{ch}^{\mathfrak{su}(n)}_{\lambda}(h)+\order{q^{k+1-\sum_i\lambda_i}}}{\prod\limits_{m=0}^{\infty}\bigg[(1-q^{m+1})^{n-1}\prod\limits_{i\neq j}^{n-1}(1-h_i\bar h_jq^{m+1})\bigg]}\ ,
\end{equation}
where $\text{ch}^{\mathfrak{su}(n)}_{\lambda}(t)$ is the finite $\mathfrak{su}(n)$ character for the representation $\lambda$, and $C^{(n)}(\lambda)$ is the quadratic Casimir (see~\eqref{defCasimir}).

Similarly, the vacuum character of $\mathfrak{su}(n+1)_k$ for large level $k$ becomes (see e.g.\ \cite{Gaberdiel:2014vca})
\begin{equation}
\begin{split}
\chi^{\mathfrak{su}(n+1),k}_{0} & (q,i_{1}(h;\xi)) \\
= \ & \frac{q^{-\frac{n(n+2)k}{24 (k+n+1)}} (1+\order{q^{k}})}{\prod\limits_{m=0}^{\infty}\bigg[(1-q^{m+1})^{n}\prod\limits_{i\neq j}^{n-1}(1-h_i\bar h_jq^{m+1})\prod\limits_{k=1}^n(1-h_k\xi^{n+1}q^{m+1})(1-\bar h_k\xi^{-(n+1)}q^{m+1})\bigg]}\ .
\end{split}
\end{equation}
Analogously the $\mathfrak{u}(1)$ character can be expanded as
\begin{equation}
\Theta_{\mu,\kappa}(q;\xi)=q^{-\frac{1}{24}+\frac{\mu^2}{2\kappa}}\frac{
\xi^{-\mu}+\order{q^{\frac{\kappa}{2}-|\mu|}}}{\prod\limits_{m=0}^{\infty}(1-q^{m+1})}
\ .
\end{equation}
Plugging the previous expansions into equation~\eqref{finite-k-decomposition-characters}, we arrive at the following large $k$ expression for the coset characters:
\begin{equation}\label{XithrougA}
\Xi^{0}_{\lambda,\mu}(q) =q^{\frac{n(n+1)}{8(k+n+1)}-\frac{C^{(n)}(\lambda)}{n+k+1}-\frac{\mu^2}{2\kappa}}\left[A_{\lambda,\mu}(q)+\order{q^{k-\sum_i\lambda_i}}+\order{q^{\frac{\kappa}{2}-|\mu|}}\right]\ ,
\end{equation}
where $A_{\lambda,\mu}$ is given by
\begin{equation}
\begin{split}
\sum_{\lambda,\mu}A_{\lambda,\mu}(q)\ \text{ch}^{\mathfrak{su}(n)}_{\lambda}(h)\ \xi^{-\mu} &= q^{-\frac{n}{12}}\frac{\theta^{\text{NS}}(q)}{\prod\limits_{m=0}^{\infty}\prod\limits_{j=1}^n(1-h_j\xi^{n+1}q^{m+1})(1-\bar h_j\xi^{-(n+1)}q^{m+1})}\\
&= \prod_{j=1}^n\left(2\sin{\frac{\theta_j}{2}}\right)\frac{\vartheta_3(q,\frac{\theta_j}{2\pi})}{\vartheta_1(q,\frac{\theta_j}{2\pi})}\ ,
\end{split}
\end{equation}
and the angles $\theta_i$ are defined in analogy with~\eqref{introduction-theta-angles}.
The branching function $A_{\lambda,\mu}$ can be isolated using the orthogonality of the finite characters, and we conclude
\begin{equation}\label{coset-characters-large-k-integral}
A_{\lambda,\mu}(q)=\frac{1}{|\T^n|}\int_{\T^n}d\mu(t)\ \text{ch}^{\mathfrak{u}(n)}_{\lambda,\mu}(t,\xi)\prod_{j=1}^n\left(2\sin{\frac{\theta_j}{2}}\right)\frac{\vartheta_3(q,\frac{\theta_j}{2\pi})}{\vartheta_1(q,\frac{\theta_j}{2\pi})}\ ,
 \end{equation}
where $\T^n$ is the Cartan torus of $U(n)$.

From~\eqref{XithrougA} we see that the limit of the coset characters $\Xi^{0}_{\lambda,\mu}$ with fixed labels $\lambda,\mu$ is simply given by $A_{\lambda,\mu}$,
\begin{equation}\label{coset-characters-large-k-final}
\lim_{k\to\infty} \Xi^{0}_{\lambda,\mu}(q)= A_{\lambda,\mu}(q) \ .
\end{equation}
This provides us with an expression for the boundary partition function~\eqref{boundpartfunc} in the limit.

\subsection{Match with the continuous orbifold}

We will now show that the boundary partition functions in the limit theory coincide with the corresponding partition functions in the orbifold model $\mathbb{C}^{n}/U(n)$.

The boundary conditions in an orbifold conformal field theory arise from superpositions of conformal boundary conditions of the parent theory that are invariant under the action of the orbifold group. A boundary condition that is by itself invariant splits into ``fractional boundary conditions''.
In a continuous orbifold the only relevant boundary conditions are these fractional boundary conditions, because they couple to twisted sectors of the orbifold, which outnumber the untwisted sector in case of continuous groups (see the discussions in~\cite{Gaberdiel:2011aa,Fredenhagen:2012bw,Restuccia:2013tba}). 
Fractional boundary conditions are labelled by irreducible representations of the orbifold group (if the original boundary condition is invariant under the full orbifold group). In our case we consider the boundary condition corresponding to a point-like brane at the origin. The boundary partition function between fractional boundary conditions labelled by the $U(n)$ multi-indices $r,r'$ then reads
\begin{equation}\label{fractional-branes-on-contorbi}
Z_{rr'}(\tilde \t)=\frac{1}{|G|}\int_{G}d\m(g)\ \text{ch}^{\mathfrak{u}(n)}_r(g)\,\text{ch}^{\mathfrak{u}(n)}_{r'}(g)\,\tr_{\Hilb_{0}} \left( U(g)\tilde q ^{L_0-\frac{n}{8}}\right)\ ,\quad G=U(n)\ ,
\end{equation}
where $\Hilb_{0}$ is the Hilbert space of the boundary spectrum of a point-like brane in $\mathbb{C}^n$, $\text{ch}^{\mathfrak{u}(n)}_r$ is the finite character of the $U(n)$ representation labelled by $r$, $U(g)$ is the action of $g\in U(n)$ on the space of states $\Hilb_{0}$, and $|G|$ is the volume of $U(n)$ measured with respect to the Haar measure $d\m(g)$.

The expression in equation~\eqref{fractional-branes-on-contorbi} can be simplified by noting that
\begin{equation}
\text{ch}^{\mathfrak{u}(n)}_r(g)\,\text{ch}^{\mathfrak{u}(n)}_{r'}(g)=\sum_{s}N^{\mathfrak{u}(n)}_{rr'}{}^s\,\text{ch}^{\mathfrak{u}(n)}_{s}(g)\ ,
\end{equation}
where $N^{\mathfrak{u}(n)}_{rr'}{}^s$ are $\mathfrak{u}(n)$ Clebsch-Gordan coefficients.
This implies that any boundary partition function can be realised as a combination of elementary $Z_{r0}(\tilde{\tau})$ amplitudes.

Every group element is conjugate to some element of the Cartan torus (quotiented by the Weyl group). Using the cyclicity of the trace we can rewrite the integral as an integral over the Cartan torus of $U(n)$ parameterised as in section~\ref{sec:KS-models} by $n$ angles $\theta_i$ as~$\text{Diag}(e^{i\theta_1},\dotsc ,e^{i\theta_{n}})$.
The trace becomes
\begin{equation}
\tr_{\Hilb_0}\left( U(g)\tilde{q}^{L_0-\frac{n}{8}}\right)=\prod_{i=1}^n\left(2\sin{\frac{\theta_i}{2}}\right)\frac{\vartheta_3(\tilde{\tau},\tfrac{\theta_i}{2\pi})}{\vartheta_1(\tilde{\tau},\tfrac{\theta_i}{2\pi})}\ .
\end{equation}
We conclude that
\begin{equation}\label{expression-boundary-spectrum-contorbi}
Z_{\cL \cM,0}(\tilde \t)=\frac{1}{|\T^n|}\int_{\T^n}d\m (t)\;\text{ch}^{\mathfrak{u}(n)}_{\cL\cM}(t)\,\prod_{i=1}^n\left(2\sin{\frac{\theta_i}{2}}\right)\frac{\vartheta_3(\tilde{\tau},\tfrac{\theta_i}{2\pi})}{\vartheta_1(\tilde{\tau},\tfrac{\theta_i}{2\pi})}\ ,
\end{equation} 
where we have labelled the $\mathfrak{u}(n)$ representation by representation labels of $\mathfrak{su}(n)$ and $\mathfrak{u}(1)$, $r\to (\cL,\cM)$.
Comparing~\eqref{expression-boundary-spectrum-contorbi} with equation~\eqref{coset-characters-large-k-final} we find
\begin{equation}\label{final-claim-boundary-comparison}
Z_{\cL \cM,0}(\tilde \t)\equiv A_{\cL,\cM} (\tilde q)\ .
\end{equation}
Hence the boundary partition functions of type-A Cardy boundary conditions on the coset coincide in the limit $k\to\infty$ with the boundary partition functions of fractional boundary conditions in the continuous orbifold.
This provides further evidence that the $k\to\infty$ limit of the Grassmannian Kazama-Suzuki models is equivalent to the continuous orbifold $\mathbb{C}^{n}/U(n)$.

\subsection{Explicit expressions for $SU(3)/U(2)$}

The boundary partition functions~\eqref{final-claim-boundary-comparison} in the limit theory have been explicitly analysed in~\cite{Fredenhagen:2012bw,Restuccia:2013tba} for the case $n=1$. 
We present here the explicit characters that appear in the boundary functions in the limit for the second simplest example, $n=2$, namely for the A-type boundary conditions of the large level limit of $\mathfrak{su}(3)/\mathfrak{u}(2)$ Grassmannian coset.

In the $\mathfrak{su}(3)/\mathfrak{u}(2)$ case the integral~\eqref{expression-boundary-spectrum-contorbi} reads
\begin{equation}\label{expression-boundary-spectrum-contorbi-su(3)}
Z_{\cL \cM,0}(\tilde \t)=\frac{1}{|\T^2|}\int_{\T^2}d\m (t)\;\text{ch}^{u(2)}_{\cL\cM}(t)\prod_{i=1}^2\left(2\sin{\frac{\theta_i}{2}}\right)\frac{\vartheta_3(\tilde{\tau},\tfrac{\theta_i}{2\pi})}{\vartheta_1(\tilde{\tau},\tfrac{\theta_i}{2\pi})}\ ,
\end{equation} 
where the angles $\theta_1$ and $\theta_2$ parameterise the Cartan torus $\T^2$.
The finite $U(2)$ character has the form
\begin{equation}
\text{ch}^{\mathfrak{u}(2)}_{\cL,\cM}=\frac{\sin(1+\cL)\frac{\theta_1-\theta_2}{2}}{\sin \frac{\theta_1-\theta_2}{2}}e^{i\cM \frac{\theta_1+\theta_2}{2}}\ .
\end{equation}
The induced measure on the Cartan torus is 
\begin{equation}
\frac{d\mu(t)}{|\T^2|}=\frac{d\theta_1d\theta_2}{8\pi^2}\left(2\sin \frac{\theta_{1}-\theta_{2}}{2} \right)^{2}\ .
\end{equation}
Using the following expansion (see e.g.\ ~\cite[appendix A]{Sugawara:2012ag})
\begin{equation}
\frac{\vartheta_3(\t,\frac{\theta_i}{2\pi})}{\vartheta_1(\t,\frac{\theta_i}{2\pi})}=-i\frac{\vartheta_3(\t,0)}{\eta^3(\t)}\sum_{n\in\Z}\frac{e^{i \theta_i\left(n+\12\right)}}{1+q^{n+\12}}\ ,
\end{equation}
we can explicitly solve the integral~\eqref{expression-boundary-spectrum-contorbi-su(3)}. We find
\begin{equation}\label{type3-characters-c=6}
A^{c=6}_{\cL,\cM}(q) = q^{-\frac{5}{2}+\frac{3}{2}\cM-\frac{1}{2}\cL}\left[\frac{\vartheta_3(\t,0)}{\eta^3(\t)}\right]^2
\frac{(1-q)^3(1+q)(1-q^{1+\cL})}
{\prod\limits_{j=0}^{2}\left(1+q^{(\frac{1}{2}-j)+\frac{\cM-\cL}{2}}\right)\left(1+q^{(j-\frac{1}{2})+\frac{\cM+\cL}{2}}\right)}\ ,
\end{equation}
recalling that $\cL+\cM$ must be even ($\cL,\cM$ are $\mathfrak{u}(2)$ representation labels). In the special case $\cL=\cM=0$ we find the vacuum character,
\begin{equation}\label{vacuum-character-c=6}
A^{c=6}_{0,0}(q)=q^{-\frac{1}{4}}\prod_{n=0}^{\infty}\frac{\left(1+q^{n+\frac 32}\right)\left(1+q^{n+\frac 32}\right)\left(1+q^{n+\frac 52}\right)^2}{\left(1-q^{n+1}\right)\left(1-q^{n+2}\right)^2\left(1-q^{n+3}\right)}=\left[\frac{\vartheta_3(\t,0)}{\eta^3(\t)}\right]^2\frac{\left(1-q^{\12}\right)^4\left(1+q\right)}{\left(1+q^{\frac 32}\right)^2}\ .
\end{equation}
The expression presented in~\eqref{vacuum-character-c=6} and~\eqref{type3-characters-c=6} are the characters of irreducible representations of the unprojected Neveu-Schwarz spectrum of the discrete boundary conditions in the large level limit of $\mathfrak{su}(3)/\mathfrak{u}(2)$ Kazama-Suzuki Grassmannian cosets.
Analogously they describe point-like fractional branes in the $\mathcal{N}=2$ supersymmetric continuous orbifold $\mathbb{C}^2/U(2)$.

The value of the conformal weight of the ground states above the vacuum can be recognised as the leading exponent of the characters of equation~\eqref{type3-characters-c=6} (when we take out the overall factor $q^{-1/4}$), and one finds~\cite{Restuccia:2013tba}
\begin{align}\label{ch8:table-leading-behav-contorbi-characters}
\begin{array}{c|c}
\text{range} & \text{leading term}\\[3pt]
\hline\hline\\[-6pt]
\cM =\cL =0 & q^{0}\\[4pt]
|\cM|\leq \cL-2 & q^{\cL-1}\\[4pt]
|\cM|=\cL>0 & q^{\cL-\frac{1}{2}}\\[4pt]
|\cM|=\cL+2 & q^{\cL+1}\\[4pt]
|\cM|>\cL+2 & q^{-\frac 52 -\frac 12(-3|\cM|+\cL)}
\end{array}
\end{align}

\section{Bulk spectrum in the limit}

We now want to explore the bulk spectrum of the limit theory. If one
keeps the representation labels $(\Lambda;\lambda,\mu)$ of a bulk
field fixed while taking the limit $k\to\infty$, the corresponding
conformal weight will tend to zero (or at least to a (half-)integer) (see
eq.\ \eqref{ch3:confweight}). If on the other hand one looks at the
complete spectrum of conformal weights of primaries, there are many
fractional weights, and in the limit the conformal weights that appear
even become dense on the positive real line, so that one expects a
continuous spectrum.

This behaviour is well-known from other limit theories \cite{Runkel:2001ng,Fredenhagen:2004cj,Fredenhagen:2007tk,Fredenhagen:2010zh,Fredenhagen:2012rb,Fredenhagen:2012bw}. To obtain the spectrum in the limit theory one has to study which fields contribute to some given conformal weight (or
better to a small neighbourhood of this conformal weight).\footnote{In
addition one should also specify the charges with respect to the other
currents in the $\cal{W}$-algebra.} 
The representations that contribute to some finite non-integer conformal weight
arise from coset representations where the labels scale with the level~$k$.
The precise analysis is complicated by the
fact that one does not know all the conformal weights in the Kazama-Suzuki
models explicitly, and in general only its fractional part can be
computed (by eq.\ \eqref{ch3:confweight}).\footnote{Only for $n=1$ one can bring all coset
fields by field identifications to some standard range for which one
can determine the conformal weights directly.} 

In~\cite{Gaberdiel:2014vca} a class of coset fields was identified whose
conformal weights, which could be computed exactly, precisely match
the conformal weights we expect for the ground states of the twisted
sector of the continuous orbifold theory that is supposed to describe
the limit theory. We take this identification as a starting point to
formulate a proposal for what happens to the complete bulk spectrum in
the limit. We can perform some checks to our proposal in the
$SU(3)/U(2)$ model. In section~\ref{sec:Modular} we will then see that our
proposal precisely matches with the modular bootstrap.

\subsection{Ground states}

We expect that in the limit the representation labels of bulk fields
have to be scaled with $k$. Following~\cite{Gaberdiel:2014vca}, we write the labels $\lambda,\mu$ of the
denominator group as\footnote{In~\cite{Gaberdiel:2014vca} the
prefactor is $k$ instead of $k+n+1$, in the limit $k\to\infty$ this
difference does not play a role.}
\begin{equation}\label{deflambdaalpha}
\lambda(\alpha) = (k+n+1) \big(\alpha_{2}-\alpha_{1},\dotsc
,\alpha_{n}-\alpha_{n-1}\big) \quad ,\quad 
\mu(\alpha) = (k+n+1) \sum_{i=1}^{n}\alpha_{i} \ .
\end{equation}
By using field identification the $\alpha_{i}$ can be brought to the
range $-\frac{1}{2}\leq \alpha_{i}\leq \frac{1}{2}$. Note also that by
definition they satisfy 
\begin{equation}
\alpha_{1}\leq \alpha_{2}\leq  \dotsb \leq  \alpha_{n} \ .
\end{equation}
For a given denominator label we expect that we have to tune the
numerator label in a precise way to obtain a finite conformal weight
in the limit. In~\cite{Gaberdiel:2014vca} it was proposed to choose
the numerator labels as
\begin{equation}\label{defLambdaalpha}
\Lambda (\alpha) = (k+n+1)\big(\alpha_{2}-\alpha_{1},\dotsc
,\alpha_{m}-\alpha_{m-1},-\alpha_{m},\alpha_{m+1},\alpha_{m+2}-\alpha_{m+1},\dotsc
,\alpha_{n}-\alpha_{n-1}\big) \ ,
\end{equation}
where the integer $m$ with $0\leq m\leq n$ is the number of negative
$\alpha_{i}$. For those labels the conformal weight can be determined
exactly, because there appears no integer shift in the formula~\eqref{ch3:confweight} (the representation $(\lambda(\alpha);\mu(\alpha))$ occurs in the decomposition of $\Lambda(\alpha)$ as can be seen from the explicit decomposition described in appendix~\ref{app:decomposition}). 
If we consider the representations
$(\Lambda(\alpha);\lambda(\alpha),\mu(\alpha))$ for which the parameters
$\alpha_{i}$ have a finite limit, the conformal weight is given by
\begin{equation}
h_{(\Lambda(\alpha);\lambda(\alpha),\mu(\alpha))} =
\frac{1}{2}\sum_{i=1}^{n}|\alpha_{i}| +\mathcal{O} (1/k)\ . 
\end{equation}
We will now formulate a proposal for the complete spectrum of coset primaries.

\subsection{Full coset spectrum in the limit}\label{sec:fullcoset}

For given denominator labels $\lambda(\alpha),\mu(\alpha)$ we have
seen that one obtains a finite conformal weight in the limit if one
chooses the numerator label as $\Lambda(\alpha)$. Except for the
selection rules the numerator labels and denominator labels run
independently, so there are many more coset fields in the
spectrum. On the other hand, as emphasised before, to get a finite
limit of the conformal weight the scaling of the labels has to be
tightly correlated, and we expect that to get a finite result the
numerator labels $\Lambda$ should deviate from $\Lambda(\alpha)$ in
the limit only by a finite amount. Taking the selection rules into
account, such $\Lambda$ are of the form
\begin{equation}\label{defLambdaalphaN}
\Lambda(\alpha,N) = \Lambda(\alpha) + \big(N_{1}-N_{2},\dotsc
,N_{m-1}-N_{m},N_{m}+N,N_{m+1}-N,N_{m+2}-N_{m+1},\dotsc ,N_{n}-N_{n-1}
\big) \ ,
\end{equation}
where $m$ again denotes the number of negative $\alpha_{i}$ and
$N_{i}$ are integers, and
\begin{equation}
N=\sum_{i=1}^{m}N_{i} - \sum_{i=m+1}^{n}N_{i} \ .
\end{equation}
The $N_{i}$ will be kept fixed in the limit. One can show (see appendix~\ref{app:decomposition}) that for
non-negative $N_{i}$ the $U(n)$ representation~$(\lambda(\alpha),\mu(\alpha))$
is contained in the decomposition of~$\Lambda(\alpha,N)$, therefore
there is no shift for the conformal weight when using the
formula~\eqref{ch3:confweight}. The result is
\begin{equation}\label{confweightpositiveN}
h= \sum_{i=1}^{n} |\alpha_{i}| \bigg(\frac{1}{2}+N_{i} \bigg) +
\mathcal{O} (1/k) \qquad (N_{i}\geq 0) \ . 
\end{equation} 
For negative $N_{i}$ we propose to shift the conformal weight by
$|N_{i}|-\frac{1}{2}$ so that the conformal weight in the limit is 
\begin{equation}\label{confweightlimit}
h= \sideset{}{^{+}}\sum_{i} |\alpha_{i}| \bigg(\frac{1}{2}+N_{i} \bigg) + 
\sideset{}{^{-}}\sum_{i} \big(1-|\alpha_{i}| \big) \bigg(-\frac{1}{2}+|N_{i}| \bigg)
+\mathcal{O} (1/k)  \ , 
\end{equation}
where the sum with superscript '$+$' runs over those $i$ for which
$N_{i}\geq 0$, and the one with the superscript '$-$' over those with $N_{i}<0$.

We cannot give a proof of the proposed shift in the conformal weight,
but we can give some justifications. First of all there has to be a shift,
because $(\lambda(\alpha),\mu(\alpha))$ does not occur in the
decomposition of $\Lambda(N,\alpha)$ if any of the $N_{i}$ is
negative (see appendix~\ref{app:decomposition}). In~\cite{Gaberdiel:2014vca} it was argued that finite
changes of $\Lambda$ away from $\Lambda(\alpha)$ would result in
representations belonging to the same twisted sector labelled by the
$\alpha_{i}$ in the orbifold description. In the twisted sector we
expect all conformal weights of primaries to be half-integer multiples of
$|\alpha_{i}|$ and $(1-|\alpha_{i}|)$, and the shift we propose is
the only choice that is consistent with these orbifold expectations.

Also let us look at a simple example. Choose all $\alpha_{i}$
negative, such that $m=n$. Furthermore we choose $N_{1}=-1$ and all
other $N_{i}=0$. Then the numerator label is 
\begin{equation}
\Lambda =\Lambda(\alpha,N) = \Big(\lambda_{1}-1,\lambda_{2},\dotsc
,\lambda_{n-1}, -\frac{1}{n}\big(\mu +|\lambda|\big)-1 \Big) \ .
\end{equation}
In its decomposition to representations of $U(n)$ one finds among
others the representation 
\begin{equation}
\big(\lambda - \mathbf{f} , \mu + (n+1) \big) \ ,
\end{equation}
where $\mathbf{f}$ is the fundamental weight of $su(n)$.
On the other hand the vector representation $\mathbf{v}$ of $so(2n)$ decomposes
into the $u(n)$ representation $(\mathbf{f},n+1)\oplus
(\bar{\mathbf{f}},-n-1)$, so that $(\lambda,\mu)$ is contained in
$\Lambda\otimes\mathbf{v}$. The formula for the conformal weight
therefore has to be shifted by the conformal weight $h(\mathbf{v})=1/2$
of the vector representation of $so(2n)_{1}$.

We give some more evidence in the next subsection in the example of
$SU(3)/U(2)$. Another justification for our proposal is provided by
the modular bootstrap analysis in section~\ref{sec:Modular}.

\subsection{The example of $SU(3)/U(2)$}

We now consider the Kazama-Suzuki model based on $SU(3)/U(2)$ (i.e.\
the case $n=2$). We consider negative $\alpha_{1},\alpha_{2}$ (i.e.\
$m=n=2$), and coset label
$\big(\Lambda(\alpha,N);\lambda(\alpha),\mu(\alpha)\big)$ with
\begin{align}
\Lambda(\alpha,N) &= \big((k+3) (\alpha_{2}-\alpha_{1}) +N_{1}-N_{2} ,
-(k+3)\alpha_{2} +N_{1}+2N_{2} \big)\\
\lambda(\alpha) &= (k+3) (\alpha_{2}-\alpha_{1}) \\
\mu(\alpha)
&= (k+3) (\alpha_{1}+\alpha_{2}) \ .
\end{align}
For non-negative $N_{i}$ the conformal weight can be computed using
the coset formula~\eqref{ch3:confweight}  without (half-)integer shift, and the result
is (compare eq.\ \eqref{confweightpositiveN})
\begin{equation}
h = |\alpha_{1}|\left(N_{1}+\frac{1}{2} \right) +
|\alpha_{2}|\left(N_{2}+\frac{1}{2} \right) + \mathcal{O} (1/k)\ .
\end{equation}
For one or both $N_{i}$ negative there have to be shifts because the
$U(2)$ representation $(\lambda(\alpha),\mu(\alpha))$ does not appear
in the decomposition of $\Lambda(\alpha,\N)$. We consider now the case
where $N_{1}<0$ and $N_{2}\geq 0$. Then we make use of the field
identification (see~\eqref{ch3:KS-identifications}) and shift the labels by the action of the simple current
\begin{align}
\Lambda(\alpha,N) &\to \Lambda^{(1)}(\alpha,N) =\big(-(k+3)\alpha_{2}+N_{1}+N_{2}, (k+3)(1+\alpha_{1})-2N_{1}-N_{2}-3 \big)\\
\lambda (\alpha) &\to \lambda^{(1)} (\alpha) = (k+3)
(1-\alpha_{2}+\alpha_{1}) -2\\
\mu (\alpha) &\to \mu^{(1)} (\alpha) = (k+3)
(\alpha_{1}+\alpha_{2}+1) \ .
\end{align}
Using the explicit branching described in appendix~\ref{app:decomposition} one can now show that $(\lambda^{(1)}(\alpha),\mu^{(1)}(\alpha))$ is
contained in the decomposition of $\Lambda^{(1)}(\alpha,N)$ if
$N_{1}<0$ and $N_{2}\geq 0$. Therefore we can use the formula for the
conformal weight~\eqref{ch3:confweight}  without any (half-)integer shift, and we
obtain
\begin{equation}
h= \big(1-|\alpha_{1}| \big) \left(|N_{1}|-\frac{1}{2} \right) +
|\alpha_{2}|\left(N_{2}+\frac{1}{2} \right) + \mathcal{O} (1/k) \ .
\end{equation}
 
\section{Modular bootstrap}\label{sec:Modular}

In the last section we formulated a proposal for the bulk spectrum of
$\mathcal{N}=2$ $W_{n+1}$-primaries in the large level limit. One way of testing it is
by comparing it against the modular bootstrap: a
boundary partition function after modular transformation results in
the overlap of boundary states, which should be expressible as a
sum/integral over the contributions of the different bulk fields. 

We consider the boundary partition function $Z_{00}(q)$ for the simplest
boundary condition labelled by $(0;0,0)$, the vacuum
representation. For this we know the expression in the large level
limit (see~\eqref{expression-boundary-spectrum-contorbi}), and we can determine its modular transformation. On the other
hand, we can also determine the modular transformation at finite level
and express it as a sum over the bulk spectrum. In the limit this has
to approach the modular transform of the limit of the boundary
partition function. From the comparison we can identify the bulk
characters in the limit theory and verify that their leading exponent
matches the expectations for the conformal weight that we formulated
in the previous section.
  
\subsection{Modular transformation of the boundary partition function}

The boundary partition function for the simplest boundary condition is
given by
\begin{equation}
Z_{00} (\tilde{\tau}) = \frac{1}{|U(n)|} \int d\mu(g)\ 
\tr_{\cH_{0}} \left( U(g) \tilde{q}^{L_{0}-\frac{n}{8}} \right)\ .
\end{equation}
The integral can be rewritten as an integral over the Cartan torus
which as in section~2 we parameterise by $\text{Diag} (e^{i\theta_{1}},\dotsc
,e^{i\theta_{n}})$. The trace is then given by
\begin{equation}
\tr_{\cH_{0}} \left( U(g) \tilde{q}^{L_{0}-\frac{n}{8}}\right) =
\big(2\sin \tfrac{\theta_{1}}{2} \big) \dotsb \big(2\sin
\tfrac{\theta_{n}}{2} \big)
\frac{\vartheta_{3} (\tilde{\tau},\tfrac{\theta_{1}}{2\pi})\dotsb
\vartheta_{3}(\tilde{\tau},\tfrac{\theta_{n}}{2\pi})}{\vartheta_{1}(\tilde{\tau},\tfrac{\theta_{1}}{2\pi})\dotsb
\vartheta_{1} (\tilde{\tau},\tfrac{\theta_{n}}{2\pi})} \ .
\end{equation}
The induced measure on the torus is (Weyl's integration formula)
\begin{equation}
\frac{1}{|U(n)|} \int d\mu(g) f(g) = \frac{1}{(2\pi)^{n}n!}\int
d\theta_{1}\dotsb d\theta_{n}
\prod_{i<j} \left(2\sin\frac{\theta_{i}-\theta_{j}}{2}\right)^{\!2} f\big(\text{Diag}
(e^{i\theta_{1}},\dotsc ,e^{i\theta_{n}})\big) \ ,
\end{equation}
where $f$ is some class function on $U(n)$. Therefore the partition
function can be written as
\begin{equation}
Z_{00} (\tilde{\tau}) = \frac{1}{(2\pi)^{n}n!}\int
d\theta_{1}\dotsb d\theta_{n}
\prod_{i<j} \left(2\sin\frac{\theta_{i}-\theta_{j}}{2}\right)^{\!2}
\prod_{i=1}^{n} \left(2\sin\frac{\theta_{i}}{2} \frac{\vartheta_{3}
(\tilde{\tau},\tfrac{\theta_{i}}{2\pi})}{\vartheta_{1}
(\tilde{\tau},\tfrac{\theta_{i}}{2\pi})} \right) \ .
\end{equation}
The modular transformation of the theta-functions is well known,
\begin{align}
\vartheta_{3}\big(\tilde{\tau},\tfrac{\theta}{2} \big) &=
\sqrt{-i\tau} \,e^{i\pi\tau\left(\frac{\theta}{2\pi}\right)^{2}}
\vartheta_{3}\big(\tau,\tau\tfrac{\theta}{2\pi} \big)\\
\vartheta_{1}\big(\tilde{\tau},\tfrac{\theta}{2} \big) &=
-i\sqrt{-i\tau}\, e^{i\pi\tau\left(\frac{\theta}{2\pi}\right)^{2}}
\vartheta_{1}\big(\tau,\tau\tfrac{\theta}{2\pi} \big)\ ,
\end{align}
where $\tilde{\tau}=-\frac{1}{\tau}$, and we obtain
\begin{equation}
Z_{00} (\tilde{\tau}) = \frac{1}{(2\pi)^{n}n!}\int
d\theta_{1}\dotsb d\theta_{n} 
\prod_{i<j} \left(2\sin\frac{\theta_{i}-\theta_{j}}{2}\right)^{\!2}
\prod_{j=1}^{n}\left(2\sin \frac{\theta_{j}}{2\pi} \,\frac{i\vartheta_{3}\big(\tau,\tau\tfrac{\theta_{j}}{2\pi} \big)}{\vartheta_{1}\big(\tau,\tau\tfrac{\theta_{j}}{2\pi} \big)}\right)
\end{equation}
Now we can expand the ratio of the theta-functions,
\begin{equation}\label{ratioofthetaexpanded}
\sin \frac{\theta}{2\pi}\,\frac{\vartheta_{3} (\tau,\tau \frac{\theta}{2\pi})}{\vartheta_{1}(\tau,\tau\frac{\theta}{2\pi})}
=-i \sin \frac{|\theta|}{2\pi}\,\frac{\vartheta_{3} (\tau,0)}{\eta^{3} (\tau)} \sum_{N\in\mathbb{Z}} 
\frac{e^{i\tau|\theta|(N+\frac{1}{2})}}{1+q^{N+\frac{1}{2}}} \quad \text{for}\ -2\pi \leq \theta\leq 2\pi \ ,
\end{equation}
to find
\begin{multline}
Z_{00} (\tilde{\tau}) 
= \int d\theta_{1}\dotsb d\theta_{n} \sum_{N_{1},\dotsc ,N_{n}\in\mathbb{Z}}
\frac{1}{(2\pi)^{n}n!} \prod_{i<j} \left(2\sin \frac{\theta_{i}-\theta_{j}}{2} \right)^{\!2}
\prod_{j=1}^{n} \left(2\sin\frac{|\theta_{j}|}{2} \right)\\
\times 
\left(\frac{\vartheta_{3}(\tau,0)}{\eta^{3}(\tau)} \right)^{n}
\prod_{j=1}^{n}\frac{q^{(N_{j}+\frac{1}{2})\frac{|\theta_{j}|}{2\pi}}}{1+q^{N_{j}+\frac{1}{2}}} \ .
\end{multline}
The integrand is invariant under permutation of the $\theta_{j}$, therefore we can assume them to be ordered and multiply the expression by $n!$. After introducing the variables $\alpha_{j}=\frac{\theta_{j}}{2\pi}$, we finally arrive at
\begin{multline}
Z_{00} (\tilde{\tau}) = \int\displaylimits_{-\frac{1}{2}\leq \alpha_{1}\leq  \dotsb \leq \alpha_{n}\leq \frac{1}{2}} \mspace{-40mu} d\alpha_{1}\dotsb d\alpha_{n} \sum_{N_{1},\dotsc ,N_{n}\in\mathbb{Z}} \ 
\prod_{i<j} \left(2\sin \pi (\alpha_{i}-\alpha_{j}) \right)^{2}
\prod_{j=1}^{n} \left( 2\sin \pi |\alpha_{j}|\right)\\
\times 
\prod_{j=1}^{n}\left(  \frac{\vartheta_{3}(\tau,0)}{\eta^{3}(\tau)}  \frac{q^{(N_{j}+\frac{1}{2})|\alpha_{j}|}}{1+q^{N_{j}+\frac{1}{2}}}\right) \ .
\label{Z00modtransformed}
\end{multline}
We want to interpret this expression as an integral/sum over the bulk sectors labelled by the $\alpha_{j}$ and $N_{j}$, 
\begin{equation}
Z_{00} (\tilde{\tau}) = \int\displaylimits_{-\frac{1}{2}\leq \alpha_{1}\leq  \dotsb \leq \alpha_{n}\leq \frac{1}{2}} \mspace{-40mu} d\alpha_{1}\dotsb d\alpha_{n} \sum_{N_{1},\dotsc ,N_{n}\in\mathbb{Z}} \ 
S_{\{\alpha ,N \}}\, \chi_{\{\alpha ,N \}} (q)\ ,
\end{equation}
where the character of the corresponding representation is given by
\begin{equation}\label{limitcharacter}
\chi_{\{\alpha ,N \}} (q) = \prod_{j=1}^{n}\left(  \frac{\vartheta_{3}(\tau,0)}{\eta^{3}(\tau)}  \frac{q^{(N_{j}+\frac{1}{2})|\alpha_{j}|}}{1+q^{N_{j}+\frac{1}{2}}}\right) \ .
\end{equation}
The coefficient that appears in front of the character,
\begin{equation}\label{defSalphaN}
S_{\{\alpha ,N \}} = \prod_{i<j} \left(2\sin \pi (\alpha_{i}-\alpha_{j}) \right)^{2}
\prod_{j=1}^{n} \left( 2\sin \pi |\alpha_{j}|\right)\ ,
\end{equation} 
is interpreted as the coefficient of the disc one-point function of the corresponding bulk field in the presence of the boundary condition $(0;0,0)$ (note that it does not depend on the $N_{i}$). We confirm this interpretation by an analysis of the limit of the boundary states overlap in the following subsection.

\subsection{Limit of boundary state overlaps}

For finite $k$, the modular S-transformation of the vacuum character reads
\begin{equation}\label{finitekmodulartransform}
\Xi^{0}_{0,0}(\tilde{q})=\sum_{\L,\l,\m} S_{(0;0,0)(\Lambda;\lambda,\mu)} \,\Xi^{\Lambda}_{\l,\m}(q)\ ,
\end{equation}
where $\L,\l,\m$ are the labels of the bulk representation, and the sum must be performed taking care of identifications and selection rules.
The modular S-matrix is given by
\begin{equation}
S_{(0;0,0) (\Lambda;\lambda,\mu)} = n (n+1) \, S_{0\Lambda}^{\mathfrak{su}(n+1)}S_{0\lambda}^{\mathfrak{su}(n)}S_{0\mu}^{\mathfrak{u}(1)} \ ,
\end{equation}
as a product of S-matrices for $\mathfrak{su}(n+1)$, $\mathfrak{su}(n)$ and $\mathfrak{u}(1)$, which read
\begin{align}
S_{0\L}^{\mathfrak{su}(n+1)}&=\frac{(k+n+1)^{-\frac{n}{2}}}{\sqrt{n+1}}\ 
\prod_{1\leq i<j\leq n+1}\left( 2\sin \pi
\frac{\sum\limits_{k=i}^{j-1}\L_k+(j-i)}{k+n+1}\right)\\
S_{0\l}^{\mathfrak{su}(n)}&=\frac{(k+n+1)^{\frac{1-n}{2}}}{\sqrt{n+1}}\ 
\prod_{1\leq i<j\leq n}\left( 2\sin \pi
\frac{\sum\limits_{k=i}^{j-1}\l_k+(j-i)}{k+n+1}\right)\\
S_{0\mu}^{\mathfrak{u}(1)} & = \frac{1}{\sqrt{n(n+1)(k+n+1)}} \ .
\end{align}
We now evaluate $S_{0\lambda}^{\mathfrak{su}(n)}$ for $\lambda=\lambda(\alpha)$ given in~\eqref{deflambdaalpha}, and we find
\begin{equation}
S_{0\lambda(\alpha)}^{\mathfrak{su}(n)} = 
\frac{(k+n+1)^{\frac{1-n}{2}}}{\sqrt{n}}
\prod_{1\leq i<j\leq n} \big(2\sin \pi (\alpha_{j}-\alpha_{i}) + \mathcal{O} (1/k)\big) \ .
\end{equation}
Similarly we evaluate $S_{0\Lambda(\alpha,N)}^{\mathfrak{su}(n+1)}$ for $\Lambda(\alpha,N)$ given in~\eqref{defLambdaalphaN}, and we arrive at
\begin{equation}
S_{0\Lambda(\alpha,N)}^{\mathfrak{su}(n+1)} = \frac{(k+n+1)^{-\frac{n}{2}}}{\sqrt{n+1}}
\prod_{1\leq i<j\leq n} \big(2\sin \pi (\alpha_{j}-\alpha_{i}) \big) 
\prod_{i=1}^{n} \big(2\sin \pi |\alpha_{i}| \big) \left(1+\mathcal{O} (1/k) \right) \ ,
\end{equation}
where the leading term is independent of the $N_{j}$. Our final expression for the S-matrix is then
\begin{equation}
S_{(0;0,0)(\Lambda(\alpha,N);\lambda(\alpha),\mu(\alpha))} = (k+n+1)^{-n} \,S_{\{\alpha , N \}} \left(1+\mathcal{O} (1/k) \right)\ ,
\end{equation}
where $S_{\{\alpha,N\}}$ is the coefficient that we obtained from the modular transformation of the limit of $Z_{00}$ (see~\eqref{defSalphaN}).
We want to reparameterise the sum over $\Lambda,\lambda$ and $\mu$ in~\eqref{finitekmodulartransform} by $\alpha_{j}$ and $N_{j}$. The sum over $\lambda$ and $\mu$ can be replaced by an integral for large $k$, and when we do a variable transformation to $\alpha_{j}$, we get a factor $n(k+n+1)^{n}$ from the variable transformation. Because of the selection rules not all combinations of $\lambda$ and $\mu$ are allowed, which reduces the integral by a factor of $1/n$, so that we find
\begin{equation}
\sum_{\substack{\lambda,\mu\\ \text{allowed}}} \longrightarrow (k+n+1)^{n} \int\displaylimits_{-\frac{1}{2}\leq \alpha_{1}\leq \dotsb \leq \alpha_{n}\leq \frac{1}{2}} \mspace{-40mu}d\alpha_{1}\dotsb d\alpha_{n} \ .
\end{equation}
We can then write down the limit of the modular transformed boundary partition function $Z_{00}$ as
\begin{align}
Z_{00} (\tilde{\tau}) &= \lim_{k\to\infty} \Xi^{0}_{0,0}(\tilde{q}) \\
&= \lim_{k\to\infty} \sum_{\L,\l,\m} S_{(0;0,0)(\Lambda;\lambda,\mu)} \,\Xi^{\Lambda}_{\l,\m}(q)\\
&= \int\displaylimits_{-\frac{1}{2}\leq \alpha_{1}\leq \dotsb \leq \alpha_{n}\leq \frac{1}{2}} \mspace{-40mu}d\alpha_{1}\dotsb d\alpha_{n} \sum_{N_{1},\dotsc ,N_{n}\in\mathbb{Z}}
S_{\{\alpha ,N \}} \,\chi_{\{\alpha ,N \}} (q) \ ,
\end{align}
where $\chi_{\{\alpha,N\}}$ is the limit of the character of the representation $(\Lambda(\alpha,N);\lambda(\alpha),\mu(\alpha))$ when $\alpha$ is kept fixed. By comparison with the result of the modular transformation of the limit of $Z_{00}$ in~\eqref{Z00modtransformed} we can read off the character $\chi_{\{\alpha,N\}}$, which confirms our identification in eq.\ \eqref{limitcharacter}. Note also that the leading exponent of the character $\chi_{\{\alpha,N\}}$ (see~\eqref{limitcharacter}),
\begin{equation}
\chi_{\{\alpha ,N \}} = 
q^{-\frac{n}{8}+ \sum^{+}_{j}|\alpha_{j}|(N_{j}+\frac{1}{2}) + \sum_{j}^{-} (1-|\alpha_{j}|) (|N_{j}|-\frac{1}{2}) }+ \dotsb  \ ,
\end{equation}
confirms the conformal weight for the representation $(\Lambda(\alpha,N);\lambda(\alpha)),\mu(\alpha)$ that we determined in~\eqref{confweightlimit}. This gives another evidence that our prescription for the shifts of the conformal weights that we used in section~\ref{sec:fullcoset} is correct.

We arrive therefore at our final result that the primary spectrum of the limit theory is labelled by the continuous parameters $\alpha_{j}$ as well as the integers $N_{j}$. The $\mathcal{N}=2$ $\mathcal{W}_{n+1}$-characters of the corresponding representations are given in~\eqref{limitcharacter}.


\appendix

\section{Decomposition of representations}\label{app:decomposition}

A representation of $\mathfrak{su}(n+1)$ labelled by the weight $\Lambda=(\Lambda_{1},\dotsc ,\Lambda_{n})$ (with Dynkin labels $\Lambda_{i}$) decomposes into representations $(\lambda,\mu)$ of $\mathfrak{su}(n)\oplus\mathfrak{u}(1)$ embedded as described in~\eqref{embedding-H-into-G} in the following way (see e.g.\ the appendix of~\cite{Gaberdiel:2014vca}),
\begin{equation}\label{decomposition}
\Lambda \to \bigoplus_{a_{1}=0}^{\Lambda_{1}}\dotsb \bigoplus_{a_{n}=0}^{\Lambda_{n}} \left(\Big(\Lambda_{1}-a_{1}+a_{2},\dotsc ,\Lambda_{n-1}-a_{n-1}+a_{n}\Big)\,,\, -|\Lambda|+ (n+1)\sum_{i}a_{i} \right)\ .
\end{equation}
Here, $|\Lambda|=\sum_{j}j\Lambda_{j}$ is the number of boxes of the corresponding Young diagram.

As an example we test whether the representation $(\lambda(\alpha),\mu(\alpha))$ (defined in~\eqref{deflambdaalpha}) is contained in the $\mathfrak{su}(n+1)$ representation $\Lambda(\alpha,N)$ given in~\eqref{defLambdaalphaN}. If it occurs on the right hand side of~\eqref{decomposition}, the corresponding numbers $a_{i}$ are determined by the following equations:
\begin{equation}\label{eqsfora}
\begin{split}
N_{1}-N_{2}-a_{1}+a_{2} &=0\\
\vdots \\
N_{m-1}-N_{m}-a_{m-1}+a_{m} &=0\\
- (k+n+1)\alpha_{m+1}+ N_{m}+N -a_{m}+a_{m+1} &=0\\
(k+n+1) (2\alpha_{m+1}-\alpha_{m+2}) +N_{m+1}-N -a_{m+1}+a_{m+2} &=0\\
(k+n+1) (-\alpha_{m+1}+2\alpha_{m+2}-\alpha_{m+3}) +N_{m+2}-N_{m+1}-a_{m+2}+a_{m+3} &=0\\
\vdots \\
(k+n+1) (-\alpha_{n-2}+2\alpha_{n-1}-\alpha_{n}) + N_{n-1}-N_{n-2} -a_{n-1}+a_{n} &=0\\
-|\Lambda(\alpha ,N)| - (k+n+1)\sum_{j}\alpha_{j}+ (n+1)\sum_{j}a_{j} &=0 \ .
\end{split}
\end{equation}
The last equation comes from comparing $\mu(\alpha)$ with the $\mathfrak{u}(1)$ entry of~\eqref{decomposition}. To proceed we evaluate $|\Lambda(\alpha ,N)|$, and we find
\begin{equation}
|\Lambda(\alpha,N)| = \left\{ \begin{array}{ll}
\displaystyle (k+n+1)\left(  -\sum_{i}\alpha_{i} + (n+1)\alpha_{n}\right) + (n+1)N_{n}& m\leq n-1\\[4mm]
\displaystyle (k+n+1)\left(  -\sum_{i}\alpha_{i} \right) + (n+1)\sum_{j}N_{j} & m=n
\end{array}\right.
\end{equation}
Let us first consider the case $m=n$. Then the last equation of~\eqref{eqsfora} becomes simply
\begin{equation}
(n+1)\sum_{j} a_{j} - (n+1)\sum_{j}N_{j} = 0 \ ,
\end{equation}
and together with the remaining equations of~\eqref{eqsfora} it follows that
\begin{equation}\label{aintermsofNfirstcase}
a_{i} = N_{i} \quad  \text{for}\ m=n \ .
\end{equation}
Let us now consider the case $m\leq n-1$. Then the last equation of~\eqref{eqsfora} implies 
\begin{equation}
\sum_{j}a_{j} - (k+n+1)\alpha_{n} - N_{n} = 0 \ .
\end{equation}
As one can check straightforwardly, the solution to this and the remaining equations of~\eqref{eqsfora} is given by
\begin{equation}
a_{1}=N_{1}\ ,\ \dotsb \ ,\ a_{m}=N_{m}\ ,\ a_{m+1} = \Lambda_{m+1}(\alpha,N) - N_{m+1}\ ,\ \dotsb \ ,\ a_{n} = \Lambda_{n}(\alpha,N)-N_{n} \ .
\end{equation}
Comparing with~\eqref{aintermsofNfirstcase} we see that this solution also applies in the case $m=n$. The condition on the coefficients $a_{j}$ is $0\leq a_{j}\leq \Lambda_{j}(\alpha,N)$. 
For large $k$ one therefore concludes that the representation $(\lambda(\alpha),\mu(\alpha))$ appears in the decomposition of $\Lambda(\alpha,N)$ precisely if $N_{i}\geq 0$.

\end{document}